\documentclass{jpconf}
\usepackage[latin1]{inputenc}
\usepackage{graphicx}
\usepackage{amstext}

\DeclareTextFontCommand{\zapf}{\fontencoding{U}\fontfamily{pzd}\selectfont}

\def\ee{$e^+e^-$}               

\def\pp{$pp$}
\def\ppbar{$p\bar p$}

\def\nbar{\bar n}            
\def\Nbar{\bar N}            
\def\nc{{\bar n_c}}          

\def\pt{p\kern -.2pt\lower 4pt\hbox{\tiny T}}    
\def\mt{m\kern -.2pt\lower 4pt\hbox{\tiny T}}    

\def\p0{P_0(\Delta y)}

\def\avg#1{\langle #1 \rangle}  

\def\NF{\mathcal{N}_{\kern -1.9pt f}}
\def\NC{\mathcal{N}_{\kern -1.7pt c}}

\begin{document}
\title{Scenarios for multiplicity distributions in \pp\ collisions in
	the TeV energy region}

\author{Roberto Ugoccioni and Alberto Giovannini}

\address{Dipartimento di Fisica Teorica, Università di Torino and
	INFN, Sezione di Torino, via Giuria 1, 10125 Torino, Italy}

\ead{roberto.ugoccioni@to.infn.it, alberto.giovannini@to.infn.it}

\begin{abstract}
Possible scenarios based on available experimental data and
phenomenological knowledge of the GeV energy region are extended to
the TeV energy region in the framework of the weighted superposition
mechanism of soft and semi-hard events. KNO scaling
violations, forward-backward multiplicity correlations, $H_q$ vs.\ $q$
oscillations and shoulder structures are discussed.
\end{abstract}

\section{Introduction}

Given the current difficulty of performing QCD calculations in the
realm of multiparticle dynamics, one possible alternative path consists
in studying  the features of hadronic final states as
obtained in collision experiments. The main motivation is the
conviction that  the complex 
structures which we observe might very well be simple at the origin, and that 
such initial simplicity manifests itself  in terms of regularities in  final 
particle multiplicity distributions (MD's).

Indeed, the Pascal regularity \cite{ARS:report} appeared very soon: that final
charged particle MD's were well described by the negative binomial
(NB), also known as Pascal, distribution was discovered first in
cosmic rays observations \cite{Cosmic} in the '60, and later confirmed
in the 5--100 GeV energy range in \pp\ collisions and up to 40 GeV in 
\ee\ annihilations, as well as in other types of collisions
(see, e.g., \cite{Giacomelli+NA22+HRS:1+EMC}.)

The NB (Pascal) distribution is a two-parameter distribution:
\begin{equation}
			P_n^{\text{(Pascal)}}(\nbar,k) = \frac{k(k+1)\cdots(k+n-1)}{n!} 
		    \frac{\nbar^n k^k}{(\nbar+k)^{n+k}}
\end{equation}
where $\nbar$ is the average multiplicity and $1/k$ measures the
deviation of the variance $D^2\equiv\avg{n^2} - \nbar^2$ 
from the Poisson shape:
\begin{equation}
	1/k + 1/\nbar = D^2/\nbar^2 .
\end{equation}
In fact, for the Poisson distribution $D^2 = \nbar$, i.e.,
$k\to\infty$, and for the geometric distribution $D^2 = \nbar +
\nbar^2$, i.e., $k=1$.

That different reactions showed the same, approximate, regularity was
worth of interpretation attempts. Probably the most successful of such
attempt has been \emph{clan structure analysis} \cite{AGLVH:0+AGLVH:4}.
`Clan' (recalling the Scottish sense of the word) refers to a 
group of particles of common ancestry:
clans are by definition independently produced in a number which
follows the Poisson MD, each clan contains
at least one particle by assumption and all correlations are exhausted 
within each clan. 
Clan ancestors, after their production, generate additional
particles via cascading, according to a logarithmic multiplicity distribution. 

The two relevant parameters in clan structure analysis
are the average number of clans, $\Nbar$, and the average 
number of particles per  clan,  $\nc$. 
They are linked to the standard
NB (Pascal) parameters $\nbar$ and $k$ by the following non-trivial relations
\begin{equation}
	\Nbar = k \ln ( 1 + \nbar/ k )  \qquad\text{and}\qquad 
	\nc = \nbar / \Nbar .
\end{equation}
Clan structure analysis reveals new interesting properties  when  applied 
to above mentioned  collisions.
In particular it is shown that in \ee\ there are more clans than 
\pp\ collisions,   whereas each clan is much smaller.
In addition, clans in central rapidity intervals are larger 
than in more peripheral intervals.
The deep inelastic case is intermediate between the previous ones:
clans are less numerous than in \ee\ annihilation
and hadronic in character, but the clan size tends to be leptonic. 


\begin{figure}
  \begin{center}
  \mbox{\includegraphics[width=0.6\textwidth]{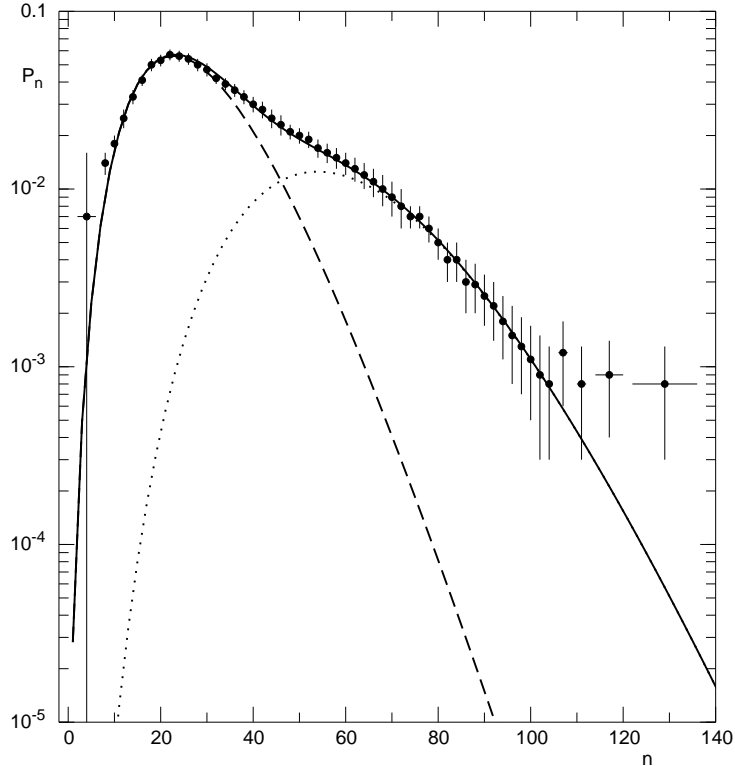}}
  \end{center}
  \caption{The shoulder structure visible in this charged particle MD
  at 900 GeV c.m.\ energy from the UA5 Collaboration \cite{Fug} 
	is well described
  by the weighted superposition of two NB (Pascal) MD (dashed and dotted
  lines).}\label{fig:fug}
\end{figure}

As soon as the c.m.\ energy increased, more detailed structures
appeared in the data: the MD was seen to have a `shoulder structure'
in the intermediate multiplicity range, first in \ppbar\ collisions
\cite{UA5:3} and later also in \ee\ annihilation \cite{DEL:1}.
In both cases an explanation was found in terms of the weighted
superposition of different classes of events: events with and without
mini-jets in the case of \ppbar\ collisions (called `semi-hard' and
`soft', respectively; see Figure~\ref{fig:fug}), 
events with a fixed number of jets
in \ee\ annihilation \cite{DEL:2}.
It should be stressed here that the MD of each class was again well
described by the Pascal distribution (which could not, of course,
by itself alone reproduce the shoulder structure). 
The degree of precision of such a description is witnessed by the
fact that not only the overall shape is visibly reproduced, but also the
quasi-oscillatory behaviour of the so-called
$H_q$ moments (defined as the ratio of factorial cumulant moments of
order $q$ to factorial moments of the same order, $H_q = K_q/F_q$),
which show sign changes when plotted as a function of the order $q$
\cite{hqlett}.

\begin{figure}
  \begin{center}
  \mbox{\includegraphics[width=0.9\textwidth]{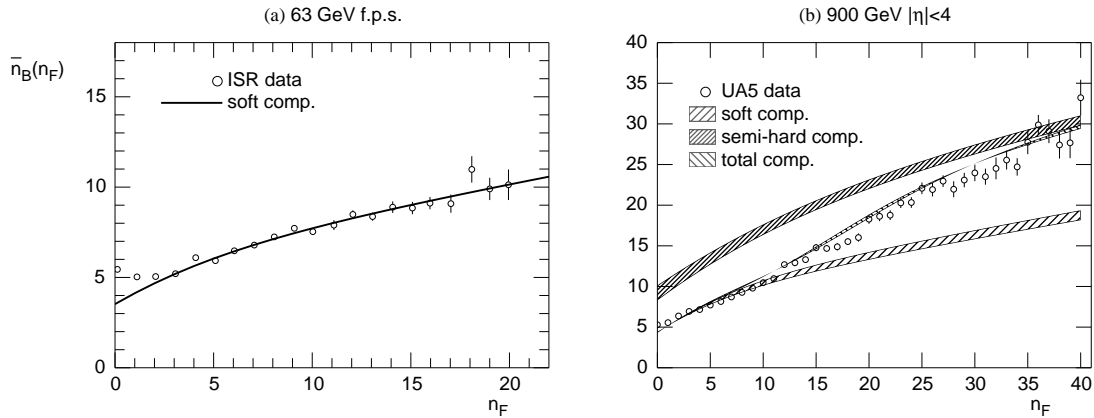}}
  \end{center}
  \caption{Results of the
  weighted superposition 
  model for $\nbar_B(n_F)$ vs.\ $n_F$ compared 
  to experimental data 
  in full phase-space at 63 GeV (a) and in the
  pseudo-rapidity interval $|\eta|<4$ at 900 GeV (b)
	\cite{RU:FB}.}\label{fig:FB}
\end{figure}

Finally, a third observable was brought into play, with different
characteristic behaviour in \ee\ annihilation and \pp\ collisions, but
which can be understood in terms of the Pascal regularity and clan
structure analysis: the strength of forward-backward multiplicity
correlations \cite{RU:FB}.
In particular, it was shown that in \ee\ annihilation the
superposition mechanism is sufficient by itself to account for the
small amount of  forward-backward multiplicity correlations (FBMC)
present in the data, consistently with the idea of a large number of
small clans; on the other hand, the superposition was not enough in
\ppbar\ collisions (Figure~\ref{fig:FB}): it is required that particle
be produced in sizeable groups (clans!) which are not localised in phase-space,
i.e., clans must produce particles in both hemispheres 
(in other words, 
there must be a non-zero `leakage': there are particles going backwards 
from forward-emitted clans, and vice versa.)

\section{Extrapolations to higher energies}

We will now abandon the parallelism with \ee\ annihilation, and 
in the following concentrate on \pp\ and \ppbar\ collisions.
The knowledge from the GeV energy range, summarised briefly in the
previous section, will now be employed to extrapolate the MD's in the
TeV energy region \cite{combo:prd,combo:eta}, 
i.e., the region covered by Tevatron and LHC colliders.

The main equation of the extrapolation is the following:
\begin{equation}
		P_n^{\text{(total)}} = \alpha_{\text{soft}}
        P_n^{\text{(Pascal)}}(\nbar_1,k_1)
    + 
		(1-\alpha_{\text{soft}}) 
        P_n^{\text{(Pascal)}}(\nbar_2,k_2) , \label{eq:3}
\end{equation}
where $\alpha_{\text{soft}}$ is the fraction of `soft' 
events in the total sample.
An event is declared `soft' if it contains no mini-jets: there is some
ambiguity as to the threshold which defines a mini-jet, e.g., a
calorimetric tower of a few GeV, but a small variation of this
threshold does not result in a wild variation of each class'
parameters. On the other hand, a proper definition of what is ``hard''
and what is ``soft'' is an extremely interesting task on itself, but
it goes well beyond the scope of this paper.

\begin{figure}
\begin{minipage}{18pc}
\includegraphics[width=18pc]{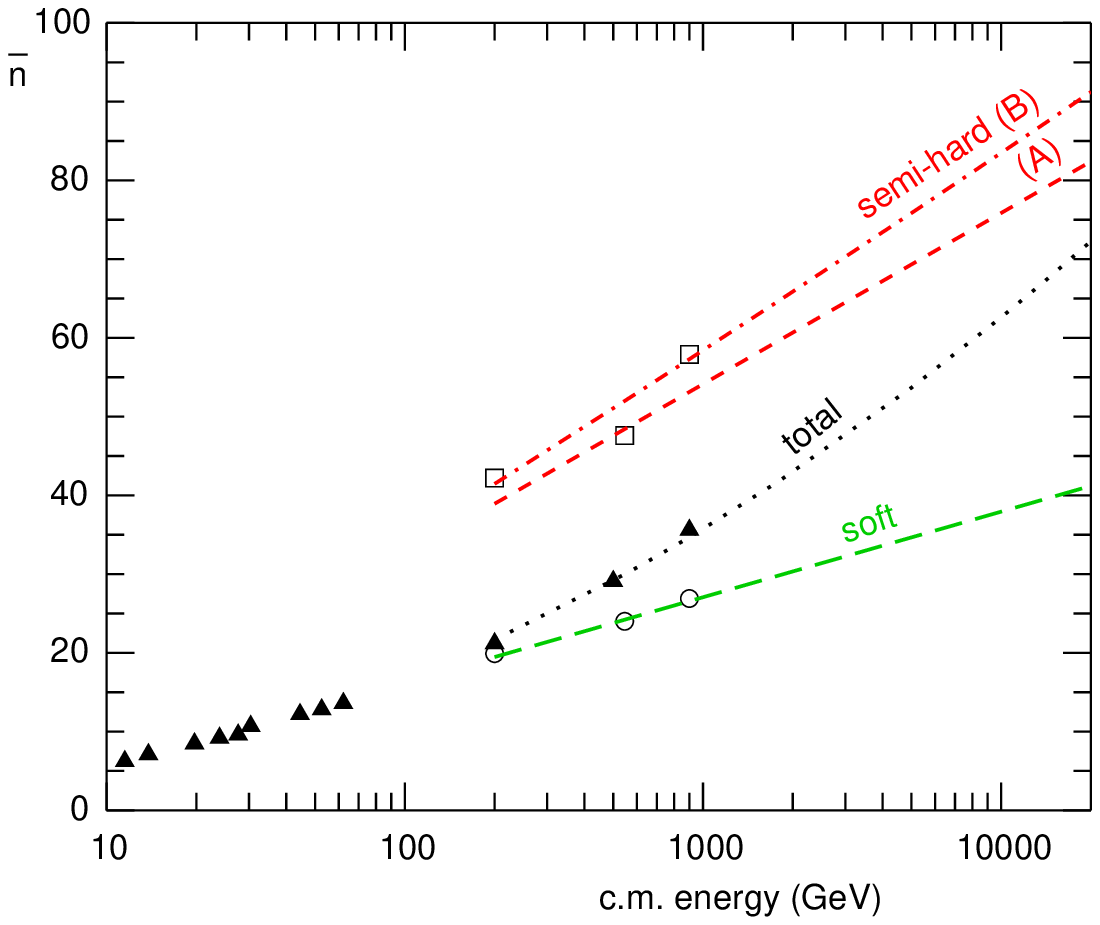}
\caption{\label{fig:extrap:n}Extrapolation of the average multiplicity.}
\end{minipage}\hspace{2pc}%
\begin{minipage}{18pc}
\includegraphics[width=18pc]{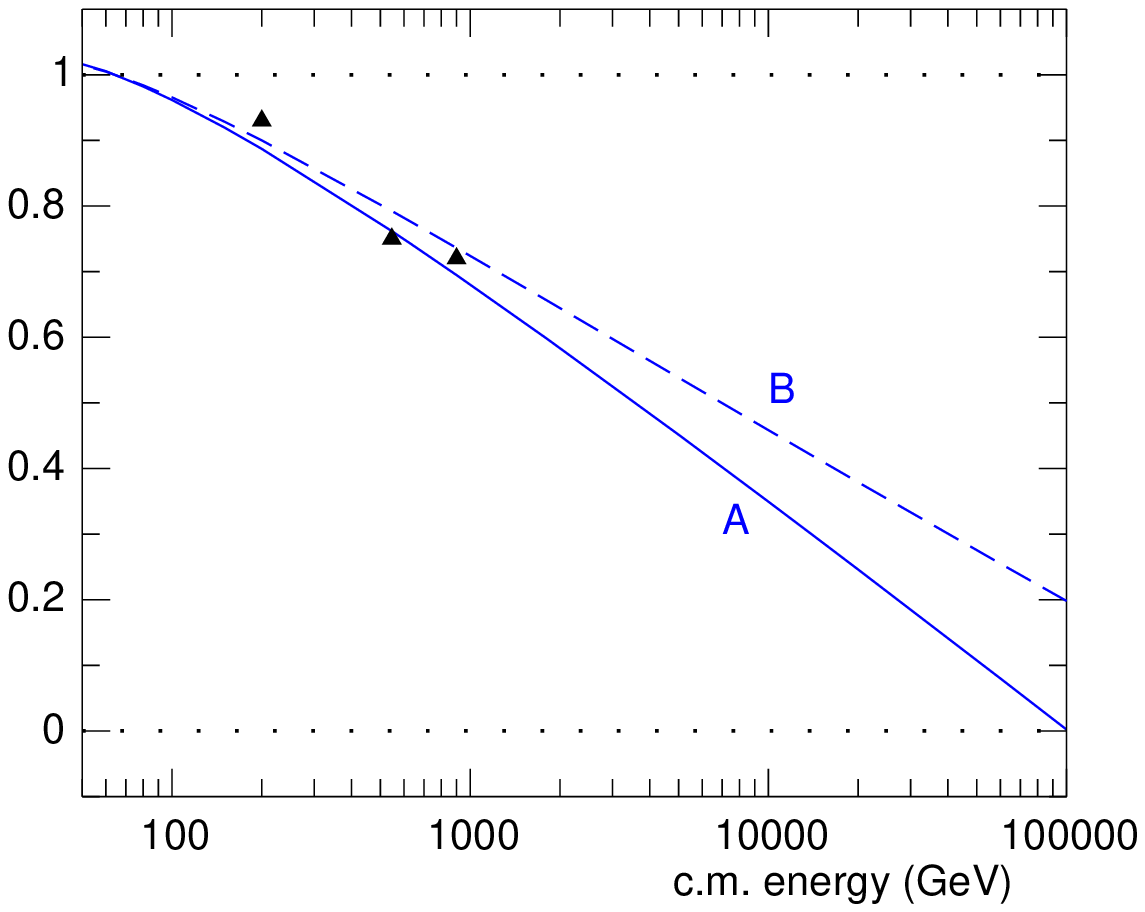}
\caption{\label{fig:extrap:a}Extrapolation of the weight parameter.}
\end{minipage} 
\end{figure}

The basic assumptions to be used in the extrapolation are as follows
(see Figure~\ref{fig:extrap:n}):

\begin{enumerate}
\item the average charged multiplicity in the total sample grows
	with the logarithm of the c.m.\ energy (best fit to available data);

\item in agreement with the findings at ISR, where one component
	gives an adequate description of the data, the growth of the average
	charged multiplicity in the \emph{soft} sample is proportional to
	$\ln s$. 

\item in agreement with UA1 findings at SpS, the average multiplicity
	in the \emph{semi-hard} sample is twice as large as in the soft one,
	but a very small term quadratic in $\ln s$ could be added as a 
	correction (the two possibilities are called A and B in the figures).
	Accordingly, most, if not all, of the $\ln^2 s$ behaviour
	of $\nbar_{\text{total}}$ comes from the increase in the mini-jet
	production cross section.
\end{enumerate}

One now has enough information to compute the expected energy dependence of
the weight $\alpha_{\text{soft}}$:
\begin{equation}
	\alpha_{\text{soft}}(s) = 2 - \nbar_{\text{total}}(s) 
	  / \nbar_{\text{soft}}(s),
\end{equation}
which is shown in Figure~\ref{fig:extrap:a}.

We come now to the behaviour of $k$.
$k_{\text{soft}}$ was found to be constant in the GeV region by the
UA5 collaboration; together with the mentioned behaviour of 
$\nbar_{\text{soft}}$ this implies that KNO scaling behaviour is valid
for the soft component in the TeV region: 
\begin{equation}
	D^2_{\text{soft}}/\nbar^2_{\text{soft}} \approx 0.14 
	\approx \text{constant}.
\end{equation}
We stay with this assumption on $k_{\text{soft}}$.

As the data on the semi-hard component are scarce, it was decided to
explore three different scenarios for the energy behaviour of
$k_{\text{semi-hard}}$, summarised, together with the
average number of clans and the average number of particles per clan,
in Figures \ref{fig:scen1}-\ref{fig:scen3}.

\newcommand{\ohmy}[2]{\hspace*{1cm}#1\hspace*{6cm}#2\hspace*{4.5cm}~}
\begin{figure}
  \begin{center}
		\ohmy{$1/k$}{$D^2/\nbar^2$}\\
  \includegraphics[width=0.8\textwidth]{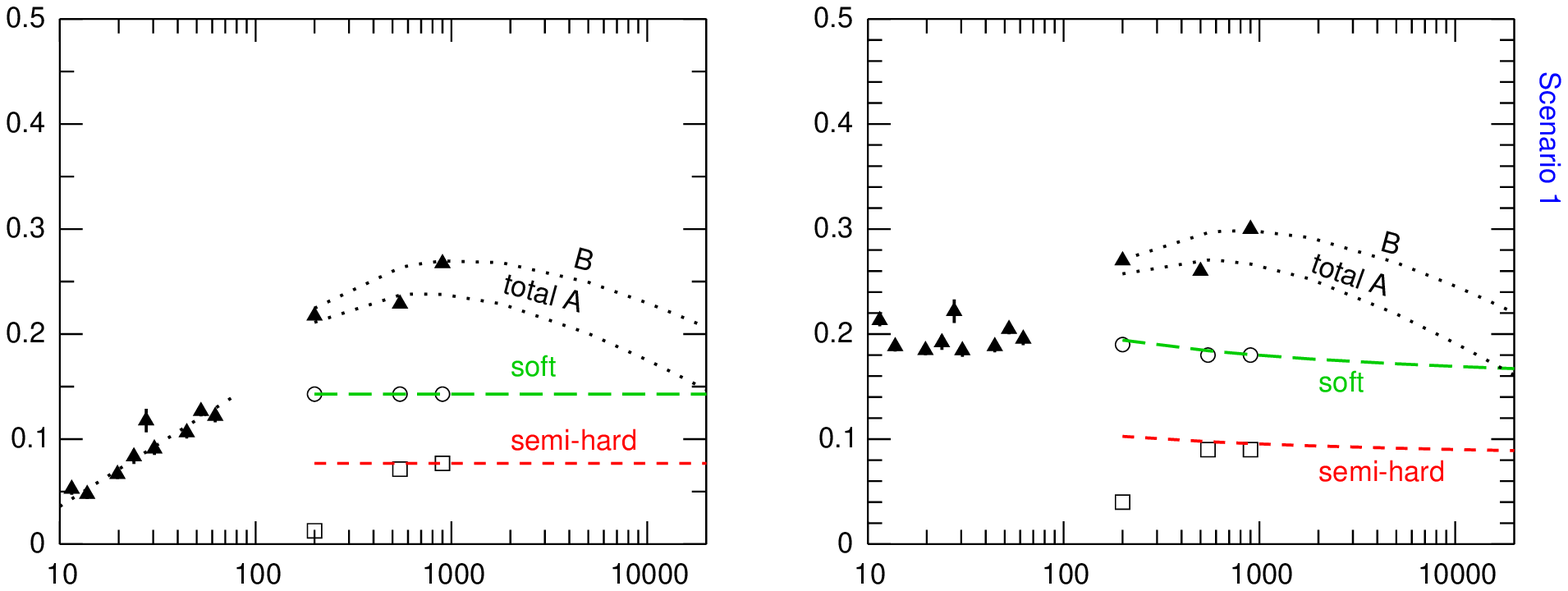}
		\ohmy{$\Nbar$}{$\nc$}\\
  \includegraphics[width=0.8\textwidth]{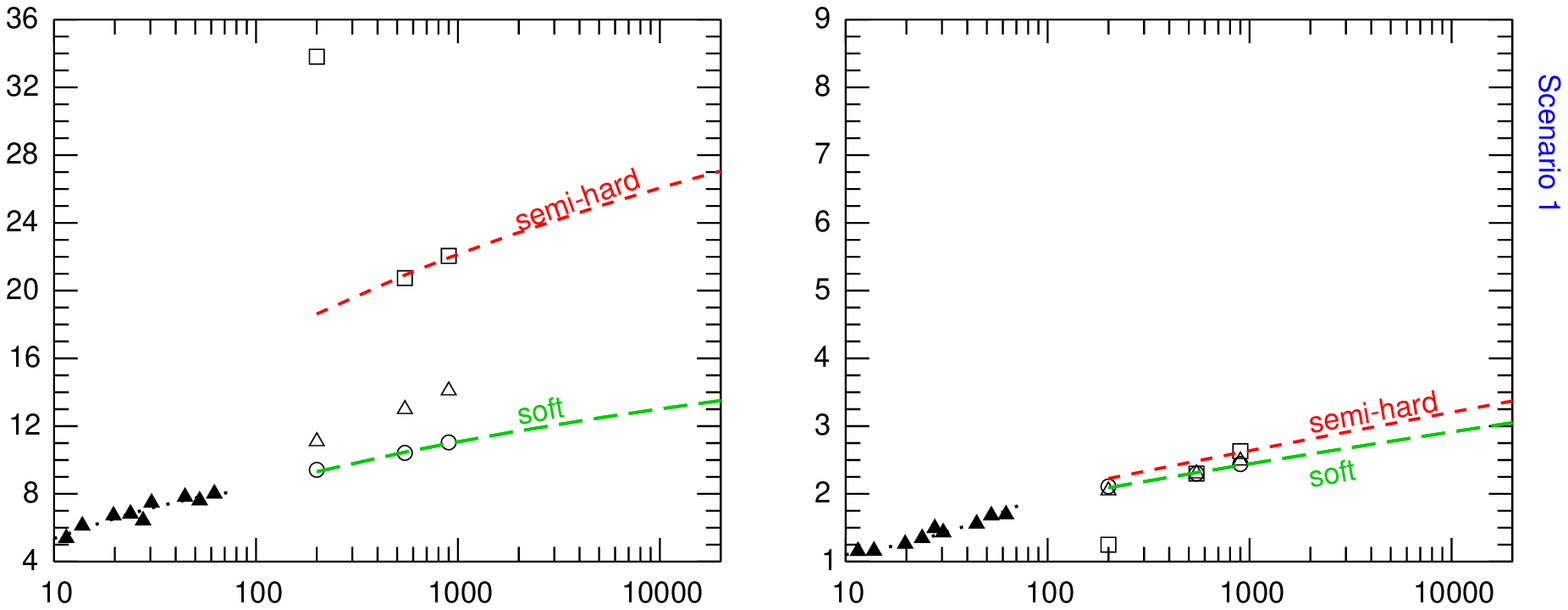}
  \end{center}
  \caption{Interpolated and extrapolated NB (Pascal) parameters in
		scenario 1.}\label{fig:scen1}
\end{figure}
\begin{figure}
  \begin{center}
		\ohmy{$1/k$}{$D^2/\nbar^2$}\\
  \includegraphics[width=0.8\textwidth]{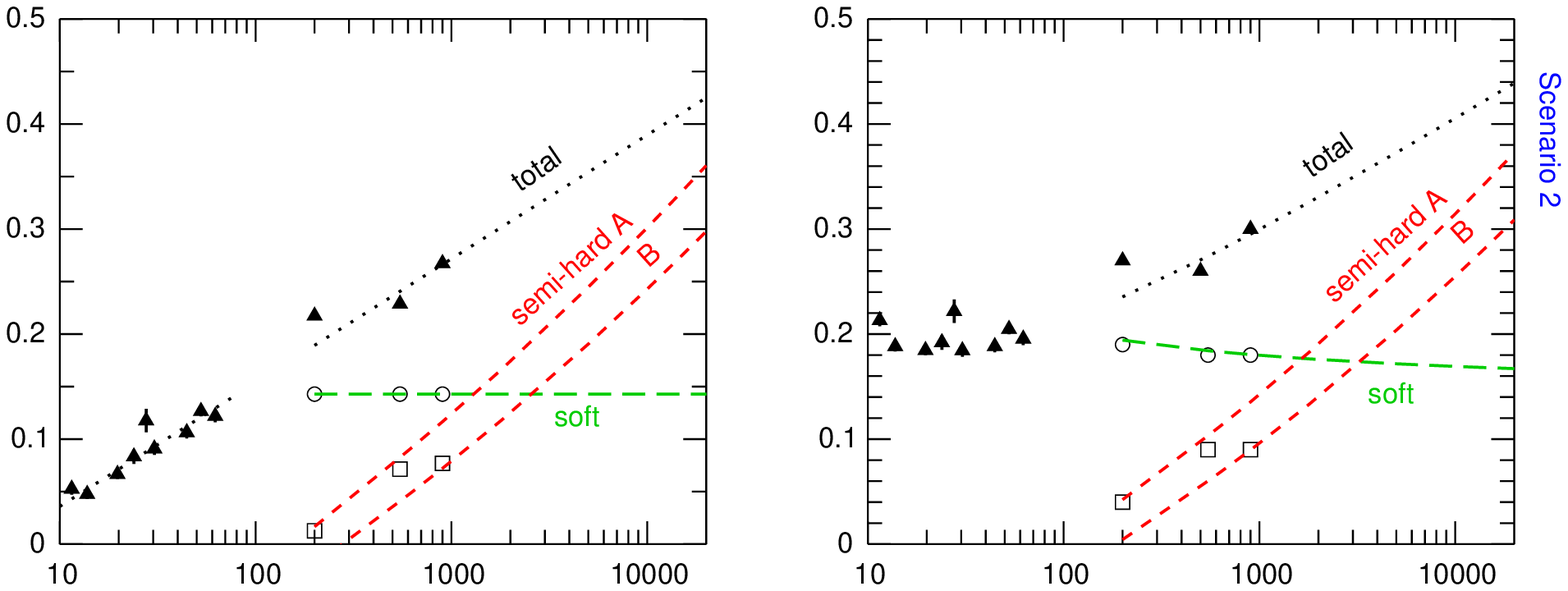}
		\ohmy{$\Nbar$}{$\nc$}\\
  \includegraphics[width=0.8\textwidth]{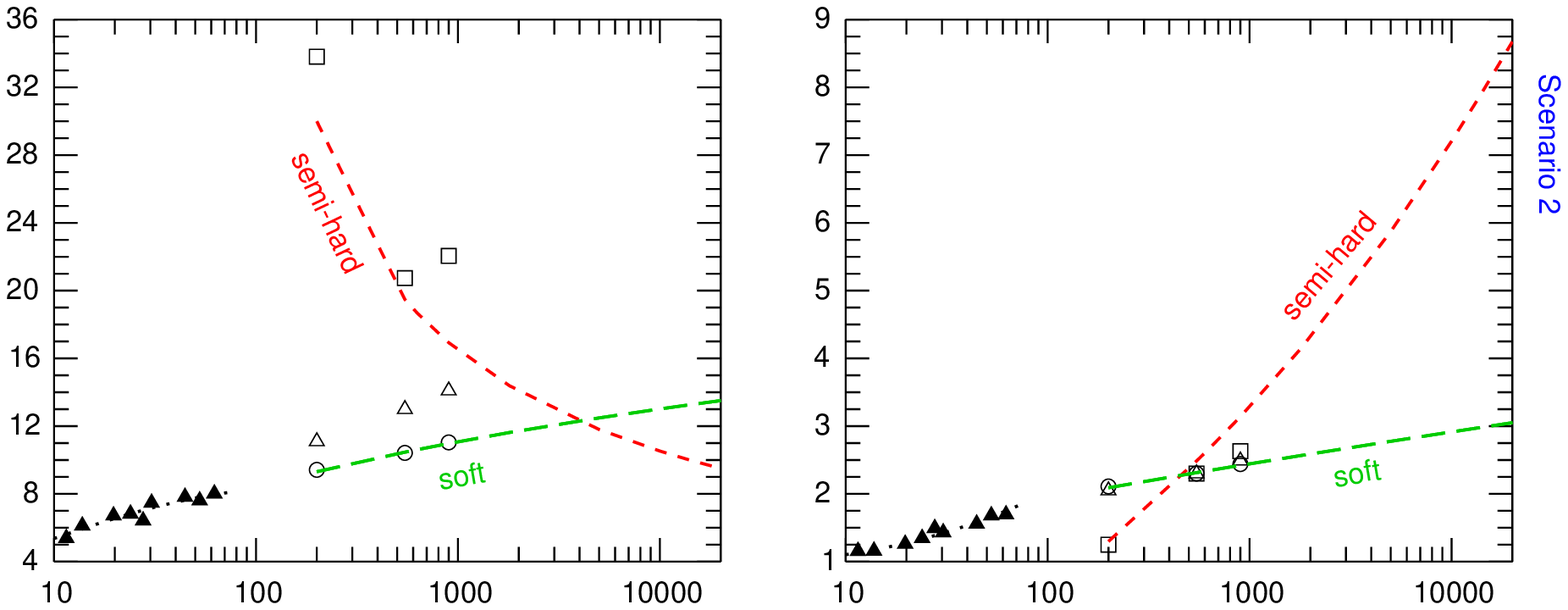}
  \end{center}
  \caption{Interpolated and extrapolated NB (Pascal) parameters in
		scenario 2.}\label{fig:scen2}
\end{figure}
\begin{figure}
  \begin{center}
		\ohmy{$1/k$}{$D^2/\nbar^2$}\\
  \includegraphics[width=0.8\textwidth]{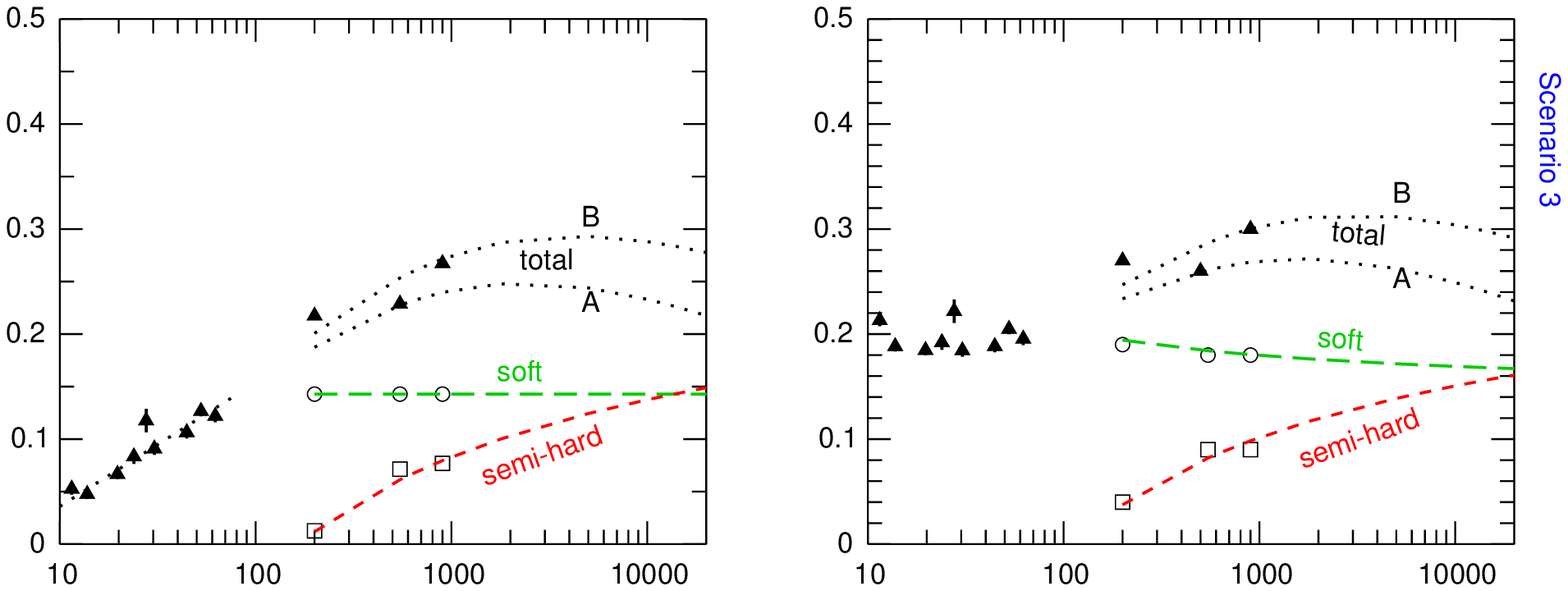}
		\ohmy{$\Nbar$}{$\nc$}\\
  \includegraphics[width=0.8\textwidth]{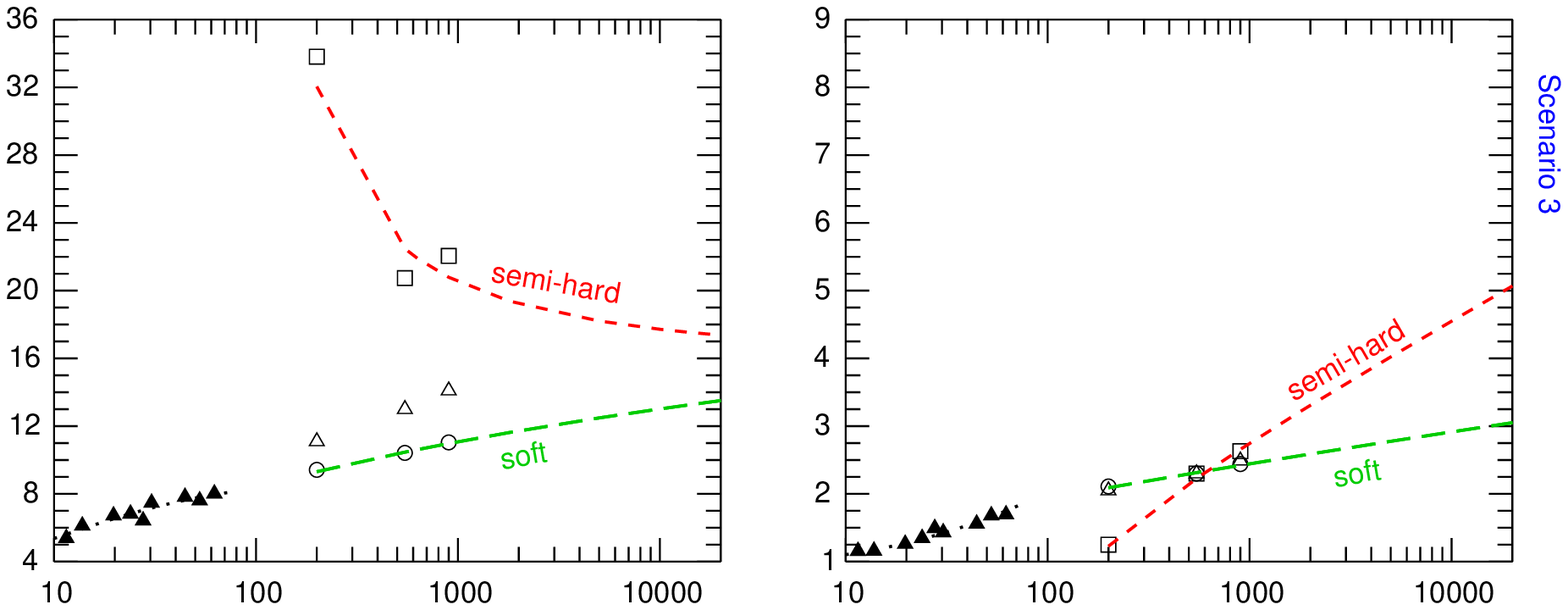}
  \end{center}
  \caption{Interpolated and extrapolated NB (Pascal) parameters in
		scenario 3.}\label{fig:scen3}
\end{figure}

\subsection*{Scenario 1 (Figure \ref{fig:scen1})}
In this scenario we assume that KNO scaling holds also for the
semi-hard component, i.e., that $k_{\text{semi-hard}}$ is constant.
This leads to a slow increase of the average number of clans and of
the average number of particles per clan.

\subsection*{Scenario 2 (Figure \ref{fig:scen2})}
Here we test the assumption of maximum KNO scaling violation, by
assuming that the growth $k_{\text{total}} \simeq 0.0512\ln
\sqrt{s} - 0.08$ continues in the TeV region. One notices that this
scenario implies a fast decrease with increasing energy of the average
number of clans, a fact which is quite unusual and interesting, and
is explored further in these proceedings
\cite{Giovannini:these}.

\subsection*{Scenario 3 (Figure \ref{fig:scen3})}
In this scenario we transport the QCD-predicted (at leading log level)
behaviour to fit the low energy \ppbar\ data: $k_{\text{semi-hard}}^{-1}
= 0.38 - \sqrt{0.42/\ln(\sqrt{s}/10)}$. It corresponds to an increase,
slower than in scenario 2, toward an asymptotic constant value, i.e.,
toward a MD which is asymptotically KNO scaling.
From the point of view of clan analysis, this scenario is intermediate
between 1 and 2, and shows again a decrease in $\Nbar$ with increasing
$\sqrt s$ ($\Nbar$ will start to increase again in the KNO scaling
regime, due to the increase of $\nbar$.)

\section{Extension to small rapidity intervals}
\begin{figure}
  \begin{center}
  \mbox{\includegraphics[width=0.98\textwidth,height=0.9\textheight]{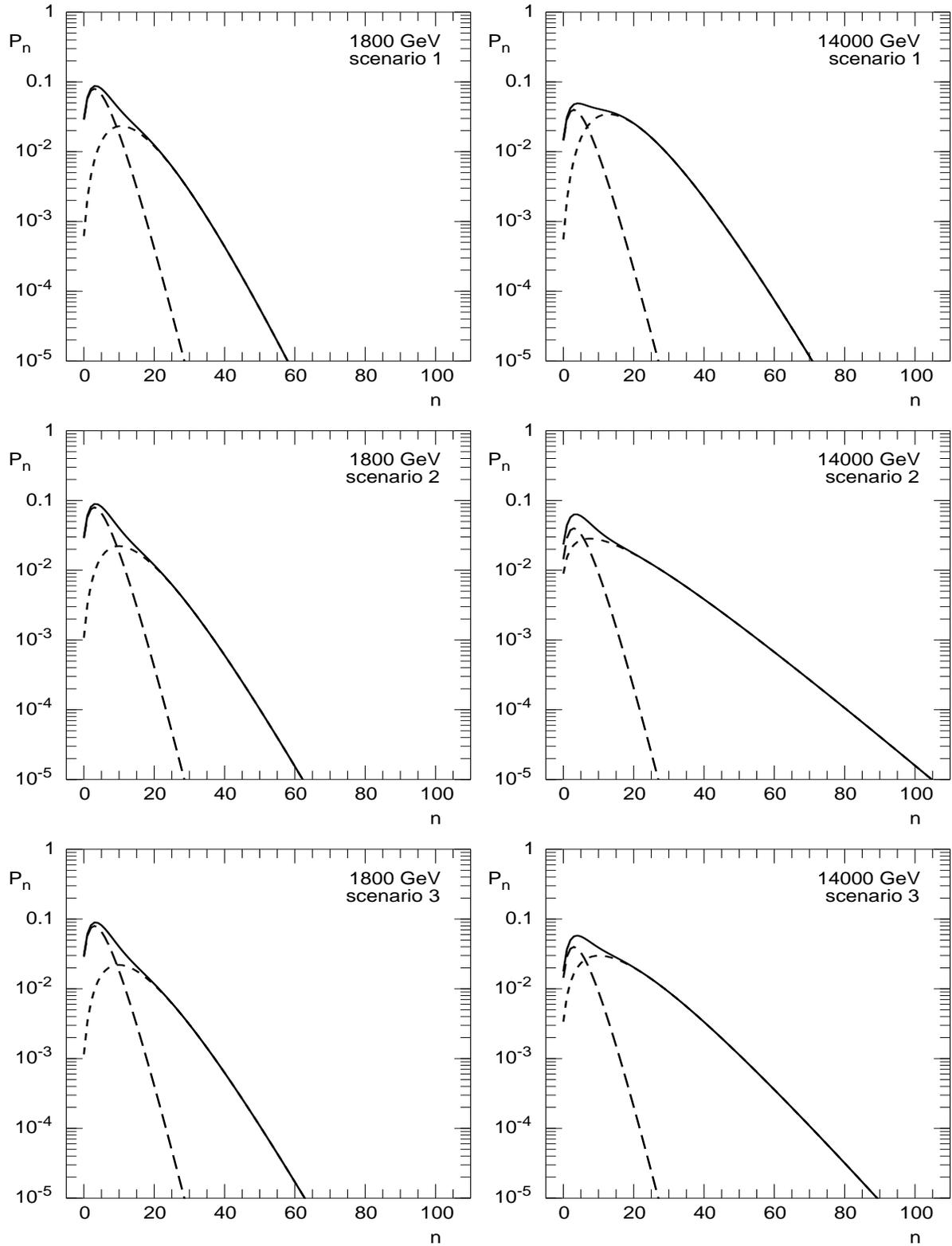}}
  \end{center}
  \caption{Predictions for the multiplicity distributions  in
		$|\eta|<1$ at 
		1800 and 14000 GeV in our scenarios for \pp\ collisions.}\label{fig:md_GU}
  \end{figure}

In going from full phase-space (FPS) to pseudo-rapidity ($\eta$) intervals,
our main concern is to be consistent with the scenarios explored in
FPS and extend them.
It should be pointed out that by assuming 
only a longitudinal growth of  phase space and constant height of the 
rapidity  plateau with c.m.\ energy for semi-hard events, as done in 
Ref.~\cite{combo:eta}, 
CDF data \cite{CDF:dNdeta} in pseudo-rapidity 
intervals  are underestimated.
These data are well
described by allowing a $\ln^2 s$ growth of the total rapidity plateau:
from this consideration one deduces a  more appropriate growth of
the semi-hard plateau height; the constraint is that
$\nbar_{\text{semi-hard}}$ 
in full phase space follows a logarithmic growth with $\sqrt{s}$ as
discussed above (curves B).
Predicted charged particle multiplicity  distributions 
in the three scenarios of the two-component model
for the interval  $|\eta| < 1$ at Tevatron and LHC energies, calculated
for the last mentioned case, are shown 
in Fig.~\ref{fig:md_GU} (notice
that a direct comparison with CDF data, in view of the relatively large
value of their resolution $\pt > 0.4$~GeV/$c$, is questionable).
If this behaviour for the semi-hard component will be confirmed  by data
one should conclude  that semi-hard events populate mainly the central 
rapidity region giving an important contribution to the increase of
charged particle density in central rapidity intervals.

\begin{figure}
  \begin{center}
  \includegraphics[width=0.45\textwidth]{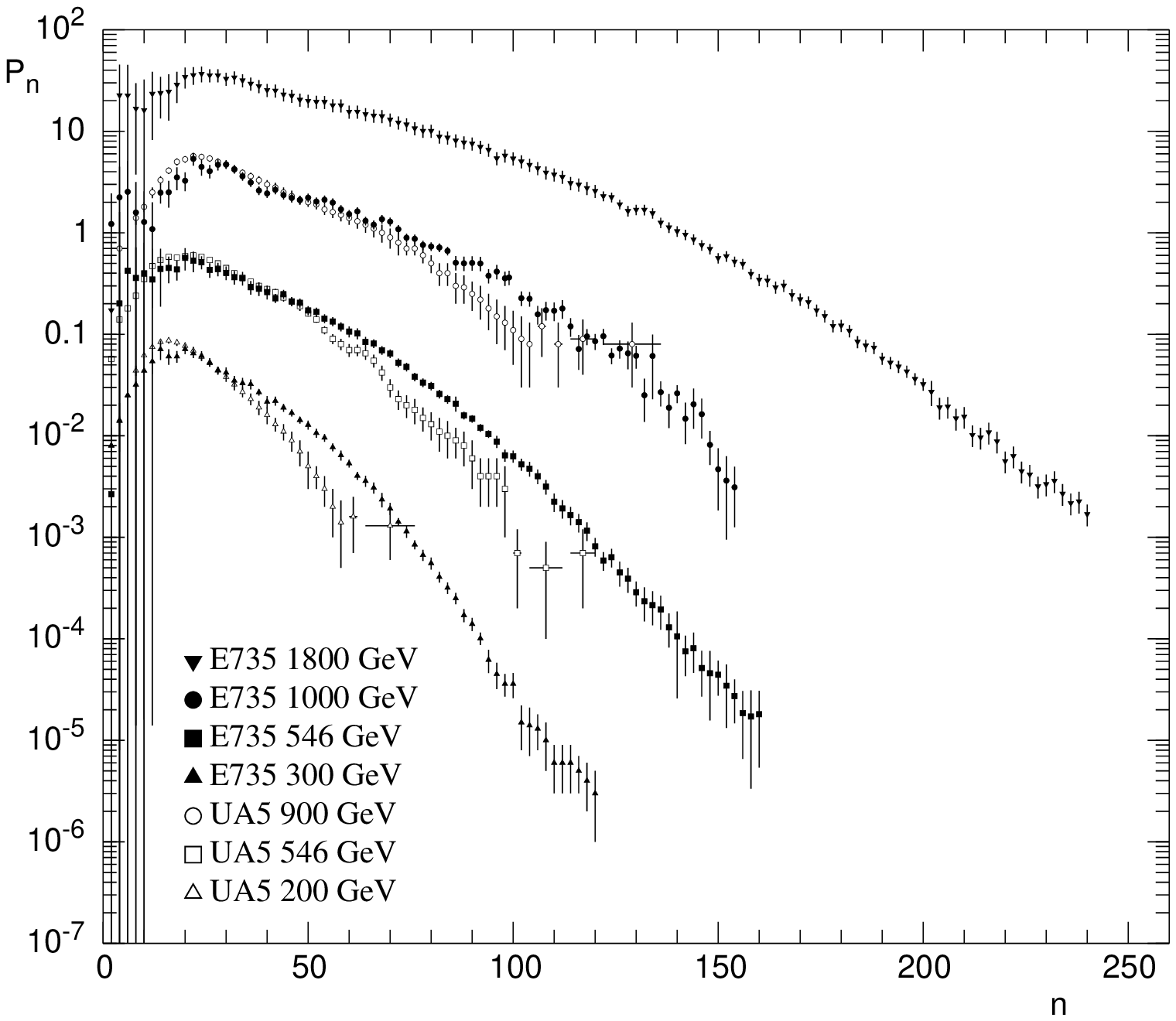}
  ~~\raisebox{0.5cm}{\includegraphics[scale=0.71]{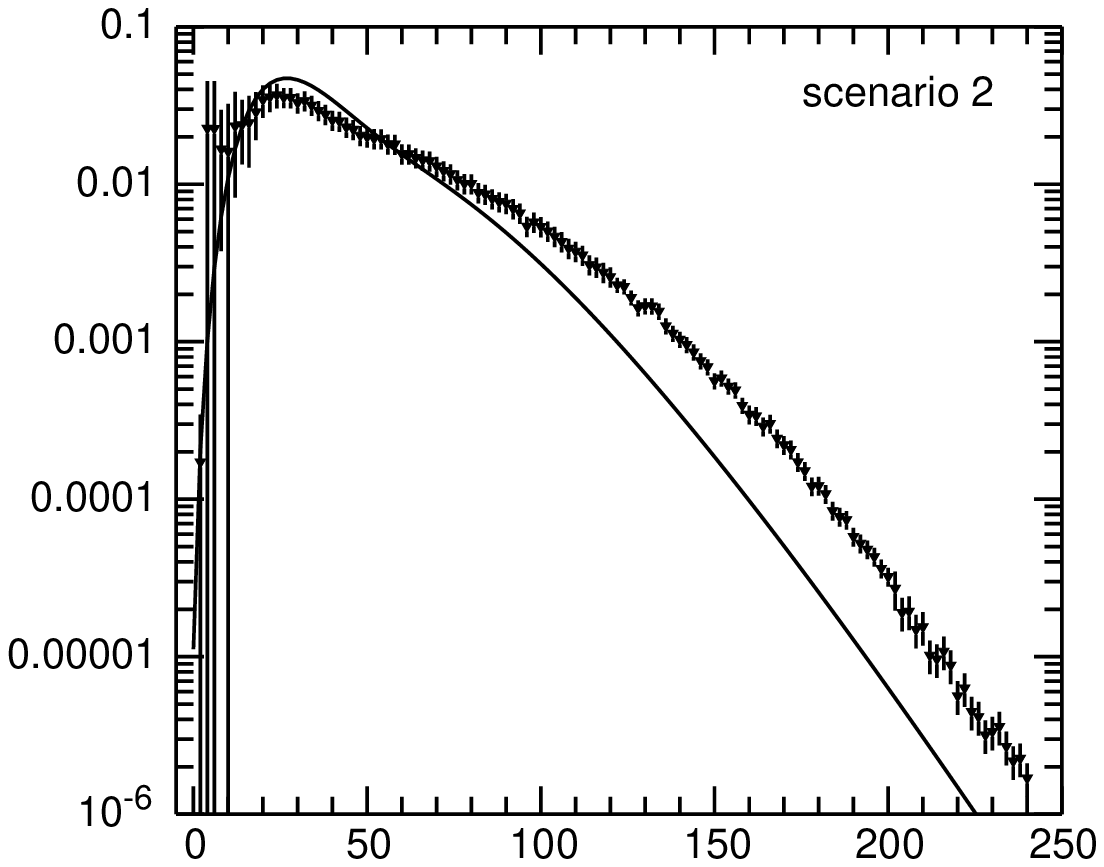}}
  \end{center}
  \caption{E735 results do not quite agree with UA5 ones (left panel;
  data from the two experiments which were taken at nearly the same
  energy are here rescaled by the same factor), 
	but are close to scenario 2's predictions (right panel).}\label{fig:e735}
\end{figure}

\begin{table}
\begin{minipage}[t]{18pc}
  \caption{NB (Pascal) and clan structure analysis parameters for
  \pp\ collisions at 14 TeV in the two-component
  model.}\label{tab:md_GU}
	\begin{center}
		\lineup
		\begin{tabular}{lllllll}
      \br
			& {FPS} &    \%   &
      $\nbar$   &   $k$ &  $\Nbar$ &  $\nc$\\
			\mr
		  & soft    &      42  &   40        &    7     &       13.3    &
		  3.0\\
			& semi-hard  &    58  &   87      &     3.7     &      11.8&
		  7.4\\
			\br
			& $ |\eta|<0.9$ &    \%   &
      $\nbar$   &   $k$ & $\Nbar$ &  $\nc$\\
			\mr
		  & soft    &42 &    \04.9     &     3.4       &   3.0      &   1.6\\
			& semi-hard  & 58 &    14  &         2.0   &       4.2  &
      3.4\\
			\br
		\end{tabular}
	\end{center}
\end{minipage}\hspace{2pc}%
\begin{minipage}[t]{18pc}
  \caption{Forward-backward multiplicity correlations strength
  parameter for
  \pp\ collisions at 14 TeV in the two-component
  model.}\label{tab:fb_GU}
	\begin{center}
		\lineup
		\begin{tabular}{llll}
      \br
    & FB corr.\ strength  & FPS & $|\eta|<0.9$\\
			\mr
    & soft             &    0.41   &     0.25   \\
    & semi-hard        &    0.51   &     0.45   \\
    & total (weighted) &    0.98   &     0.92   \\
			\br
		\end{tabular}
	
\end{center}
\end{minipage}
\end{table}

\section{Fermilab results}

Two sets of data are available from Tevatron: the E735 Collaboration
\cite{Walker}
gives full phase-space results, from data measured
in $|\eta|<3.25$ and $\pt > 0.2$~GeV/$c$ and 
then extended via a Monte Carlo program, which do not completely agree
with those obtained at comparable energies 
at the SpS collider, as shown in Figure~\ref{fig:e735} (left panel); 
Tevatron data are more precise than
SpS data at larger multiplicities (they have larger
statistics and extend to larger multiplicities than UA5 data), but
much less precise at low multiplicity.
Both sets of data show a shoulder structure, but the Tevatron MD is
somewhat wider. As seen from the same Figure~\ref{fig:e735} (right
panel), scenario 2 is the one that come closer to these data.

The second set of data comes from CDF \cite{CDF:soft-hard} for
charged particles in $|\eta|<1$ and $\pt > 0.4$~GeV/$c$:
it was found
that by subdividing the minimum bias
sample into two groups, characterised respectively by the absence
(`soft' events) or the presence (`hard' events) of mini-jets,
interesting features of the reaction can be investigated.
More precisely, a `hard' event has been defined as an event with at least
one calorimeter cluster in $|\eta|<2.4$, a cluster being defined as a
seed calorimeter tower with at least 1~GeV transverse energy $E_t$ plus at
least one contiguous tower with $E_t \ge 0.1$~GeV.  
A subdivision which is interesting \emph{per se} and can be tested at 14~TeV.
In summary, the soft component is found to satisfy KNO scaling
(as expected in our scenarios), while
the hard one does not (in disagreement with scenario 1, but in
agreement with scenarios 2 and 3); also the $\avg{\pt}$ distribution scales at
fixed multiplicity in the soft component and not in the hard one;
the dispersion of $\avg{\pt}$ vs.\ the inverse of the multiplicity is
compatible with an extrapolation to 0 as $n\to\infty$ in the soft
component but not in the hard one, indicating 
in this limit a lack of correlations in
the soft component.

\begin{figure}
\begin{minipage}{18pc}
\includegraphics[width=18pc]{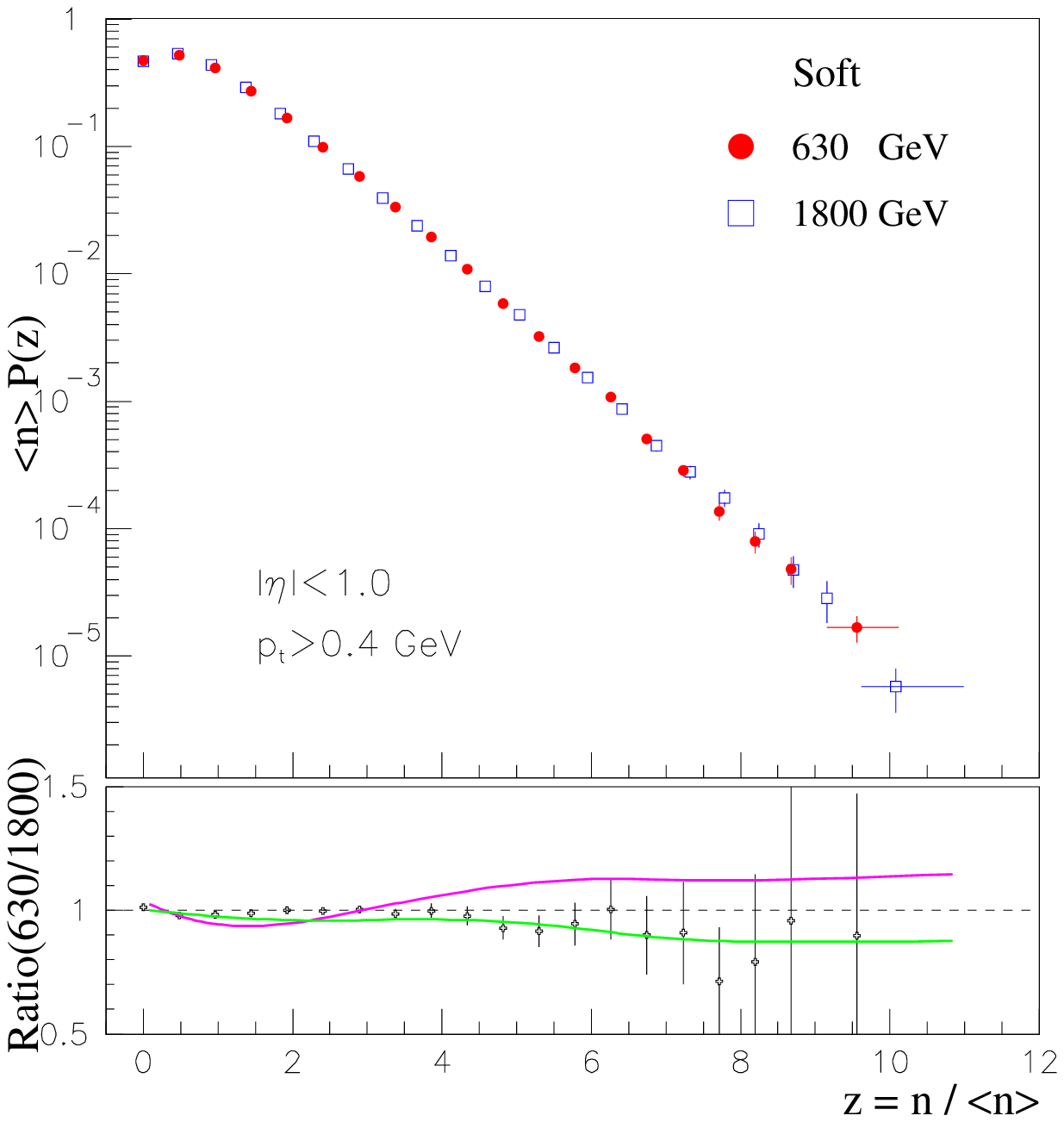}
\end{minipage}\hspace{2pc}%
\begin{minipage}{18pc}
\includegraphics[width=18pc]{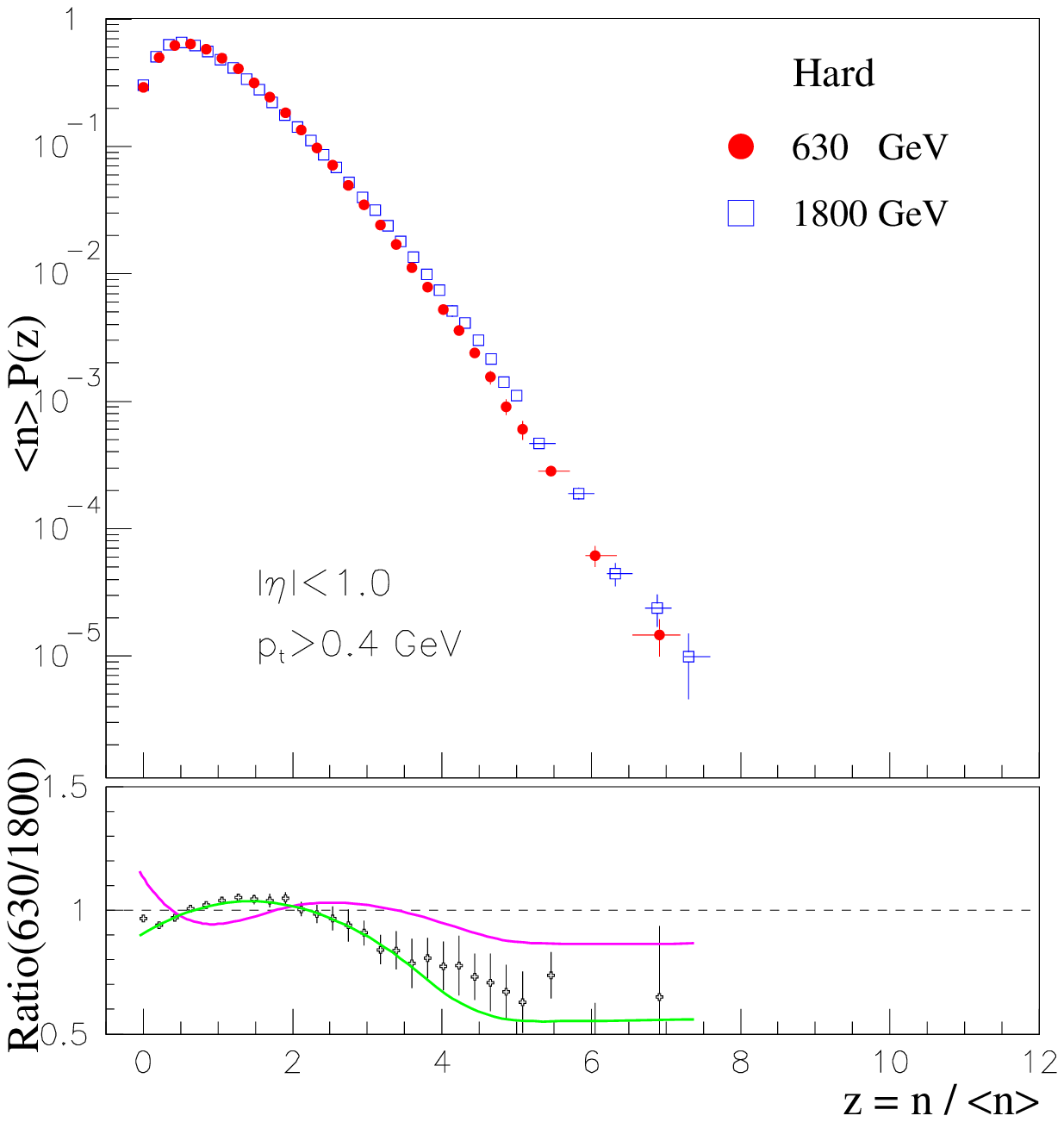}
\end{minipage}
\caption{\label{fig:cdf}CDF results on MD's at 1800 GeV: the soft
	component satisfies KNO scaling, the hard one does not \cite{CDF:soft-hard}.}
\end{figure}

\section{Conclusions}

The weighted superposition mechanism of two NB (Pascal) MD's describes
not only MD and $H_q$ oscillations, but also forward-backward
multiplicity correlations in \pp\ and \ppbar\ collisions and \ee\
annihilations. 
In hadronic collisions, the two components were found to correspond to
soft events (without mini-jets) and to semi-hard events (with
mini-jets) respectively.
Based on this mechanism, the knowledge of the features of MD's up to
900 GeV c.m.\ energy has been used to predict the characteristic
behaviour expected in the TeV energy range: the soft component
satisfies KNO scaling, while the semi-hard one violates it strongly.
These predictions do not disagree with the successive findings at
Tevatron.
In terms of clan structure analysis, a peculiar
behaviour was found: particles in the semi-hard component tend to aggregate
forming few, large clans. This behaviour, and its consequences for the
LHC experiments, will be investigated further in A. Giovannini's
contribution to these proceedings \cite{Giovannini:these}.

\section*{References}

\end{document}